\documentclass[10pt]{iopart}
\expandafter\let\csname equation*\endcsname\relax
\expandafter\let\csname endequation*\endcsname\relax 

\usepackage{xstring}
\usepackage{amsmath}
\usepackage{graphicx}
\usepackage{dcolumn}
\usepackage{bm}
\usepackage{hyperref}
\usepackage[draft]{pgf}
\usepackage{wasysym}
\usepackage{amssymb}
\usepackage{siunitx}
\DeclareSIUnit\gauss{G}
\usepackage{tensor}
\usepackage{mathrsfs}  
\usepackage{dsfont}
\usepackage{mathtools}
\usepackage{xcolor}
\usepackage{booktabs,multirow}
\usepackage{xcolor}
\usepackage{cite}
\usepackage{epstopdf}
\usepackage{braket}
\usepackage{floatrow}
\newfloatcommand{capbtabbox}{table}[][\FBwidth]

\definecolor{darkCyan}{rgb}{0, 0.5, 0.5}
\definecolor{lightGrey}{rgb}{0.5, 0.5, 0.5}

\newcommand{\refeqn}[1]{equation~\eqref{#1}}
\newcommand{\needsRef}[1]{\textcolor{darkCyan}{[?]}}
\newcommand{\chk}[1]{\textcolor{darkCyan}{\textbf{*}}}

\newcommand{\Rb}[1]{$^{#1}$Rb}
\newcommand{\Fo}{$F=1$}
\newcommand{\Ft}{$F=2$}

\newcommand{\rf}{RF}
\newcommand{\mw}{MW}
\newcommand{\gF}{\ensuremath{g_F}}

\newcommand{\Bohrmag}{\ensuremath{\mu_\mathrm{B}}}

\newcommand{\mFt}[1]{\ensuremath{\tilde{m}_F=#1}}

\begin{document}

\title[Realising a species-selective double well with MRF-dressed potentials]{Realising a species-selective double well with multiple-radiofrequency-dressed potentials}

\author{A.~J.~Barker, S.~Sunami, D.~Garrick, A.~Beregi, K.~Luksch, E.~Bentine and C.~J.~Foot}

\address{Clarendon Laboratory, University of Oxford,
	Oxford OX1 3PU, United Kingdom}
\ead{christopher.foot@physics.ox.ac.uk}

\begin{indented}
	\item[]\today
\end{indented}

\begin{abstract}

Techniques to manipulate the individual constituents of an ultracold mixture are key to investigating impurity physics.
In this work, we confine a mixture of hyperfine ground states of $^{87}$Rb atoms in a double-well potential. 
The potential is produced by dressing the atoms with multiple radiofrequencies.
The amplitude and phase of each frequency component of the dressing field are individually controlled to independently manipulate each species.
Furthermore, we verify that our mixture of hyperfine states is collisionally stable, with no observable inelastic loss.

\end{abstract}

\vspace{2pc}


\section{Introduction}

Cold atom experiments have emerged as a valuable tool to engineer many-body quantum systems~\cite{Bloch2008,Bloch2012}. 
For instance, mixtures of cold atoms have been used to study the superfluid properties of bosons and fermions~\cite{Buchler2003,DeSalvo2019,Schreck2001,Crane2000,Ferrier-Barbut2014}, and the immiscibility of quantum fluids~\cite{Papp2008,Lee2016}.
Impurity physics, in which a minority species interacts with a large reservoir of a second species, can be studied by immersing probes into a larger quantum system, which necessitates a means of control for separate constituents of the mixture~\cite{Usui2018}. 
Furthermore, the species-selective control of mixtures in double-well potentials can be used to probe several physical effects, including the excitation spectrum of a Bose-Einstein condensate (BEC)~\cite{Hangleiter2015}, the decoherence of impurities coupled to open quantum systems~\cite{Cirone2009} and to realise \mbox{sub-\si{\nano\kelvin}} thermometry techniques~\cite{Mehboudi2019,Sabin2014,Correa2015}.
Many optical methods of species-selective control have been investigated, including optical tweezers~\cite{Onofrio2002,Catani2012}, holographic light fields and lattices~\cite{Kuhn2014,Bonnin2013,Leblanc2007}.
These methods are well established but are unsuitable when the optical frequencies required to manipulate the separate components are similar, where scattering induces significant heating of the trapped mixture.

Atoms in a static magnetic field can be confined by the application of a strong radio-frequency (RF) field, forming an \rf{}-dressed potential~\cite{Hofferberth2006,Perrin2017}.
In contrast to optical traps, these potentials are smooth, have low heating rates and are free from defects; these features make them well suited to investigations of out-of-equilbrium phenomena~\cite{Rauer2017,DAlessio2015,Gring2012,Schumm2005}. 
For specific choices of magnetic and \rf{} fields, these potentials can also reduce the effective dimensionality of a trapped gas to one or two dimensions~\cite{Zobay2004,Merloti2013,Schumm2005}.
Additionally, irradiating atoms with multiple \rf{} (MRF) fields significantly extends the range of possible trapping geometries, to include double-well potentials~\cite{Harte2018,Luksch2019}, ring traps~\cite{Morizot2006,Polo2019}, toroids~\cite{Heathcote2008,Fernholz2007,Chakraborty2017} and lattices~\cite{Sinuco-Leon2015}.

\rf{}-dressed potentials are species-selective for mixtures of atoms with differing magnetic moments, for instance where the magnitude or sign of the Landé g-factor for each constituent is distinct~\cite{Bentine2017,Bentine2019a,Extavour2006,Mas2019}.
Exploiting this feature, we implement a species-selective double well using MRF-dressed potentials and manipulate the spatial distribution of the individual mixture constituents.
Recent theoretical~\cite{Owens2017} and experimental~\cite{Bentine2019a} studies have shown that not all mixtures are collisionally stable when confined in \rf{}-dressed potentials. 
In this work, we demonstrate the merit of an \rf{}-dressed mixture of \Rb{87} in the \Fo{} and \Ft{} hyperfine ground states, which we find to be long-lived.

This paper is structured as follows. In section~\ref{sec:theory} we explain the dressed-atom picture and \rf{}-dressed potentials as a method of confinement for ultracold atoms. 
In section~\ref{sec:methods} we outline the experimental procedure used to produce ultracold mixtures of hyperfine states of \Rb{87}.
In section~\ref{sec:doublewell} we present species-selective manipulations of a mixture of \Rb{87} \Fo{} and \Ft{} via control of the MRF-dressing fields amplitudes and polarisations.
In section~\ref{sec:collisions} we show that this mixture is collisionally stable against inelastic loss.

\section{Potentials of \rf{}-dressed atoms}
\label{sec:theory}

In the low-field regime, atoms in a static magnetic field $\mathbf{B}(\mathbf{r})$ have eigenenergies $m_F \gF \Bohrmag B(\mathbf{r})$, corresponding to the Zeeman substates $\ket{m_F}$ of an eigenstate with total angular momentum quantum number $F$, where \Bohrmag{} is the Bohr magneton, \gF{} is the Landé g-factor, $B(\mathbf{r}) = |\mathbf{B}(\mathbf{r})|$ and $\mathbf{r}$ is the position~\cite{Foot2005,BreitRabi}. The weak-field seeking states, for which $\gF m_F > 0$, can be trapped at local minima of $B(\mathbf{r})$.
For example, in the magnetic quadrupole field
\begin{equation}
\label{eq:quadrupole}
\mathbf{B}\left(x,y,z\right) = B'(x\mathbf{\hat{e}}_x + y \mathbf{\hat{e}}_y - 2 z \mathbf{\hat{e}}_z) \hspace{5pt} ,
\end{equation}

\noindent with field gradient $B'$ and Cartesian basis vectors \{$\mathbf{\hat{e}}_x,\mathbf{\hat{e}}_y,\mathbf{\hat{e}}_z$\}, atoms are confined around the node at the origin.

When an atom is irradiated by a strong photon field, the eigenstates of the system are conveniently described using the dressed-atom formalism~\cite{Cohen-Tannoudji1998}.
The eigenenergies depend intrinsically on the magnetic nature of the atom, via the sign and magnitude of \gF{}. 
In particular, $\mathrm{sgn}(\gF{})$ determines the handedness of the circularly polarised \rf{} field which couples the Zeeman substates: atoms with negative \gF{} couple to the $\sigma^{+}$ components of the \rf{} field, and those with positive \gF{} couple to the $\sigma^{-}$ components.
If a mixture contains atoms of differing $\mathrm{sgn}(\gF{})$, each species couples to different polarisation components in an applied \rf{} field, providing a means of manipulating each species separately.
For instance, a mixture of the hyperfine ground states of \Rb{87} with \Fo{} and \Ft{}, which have $\gF = -1/2$ and $1/2$, respectively, can be independently manipulated by controlling the amplitudes of the $\sigma^{-}$ and $\sigma^{+}$ components of the applied \rf{} field.
The interaction between the atom and the \rf{} field couples states within manifolds of constant $\tilde{N} = \mathrm{sgn}\left(\gF{}\right)m_F + N$, where $N$ indicates the Fock state $\ket{N}$ of the \rf{} field.

\begin{figure}
	\includegraphics[width=0.95\linewidth]{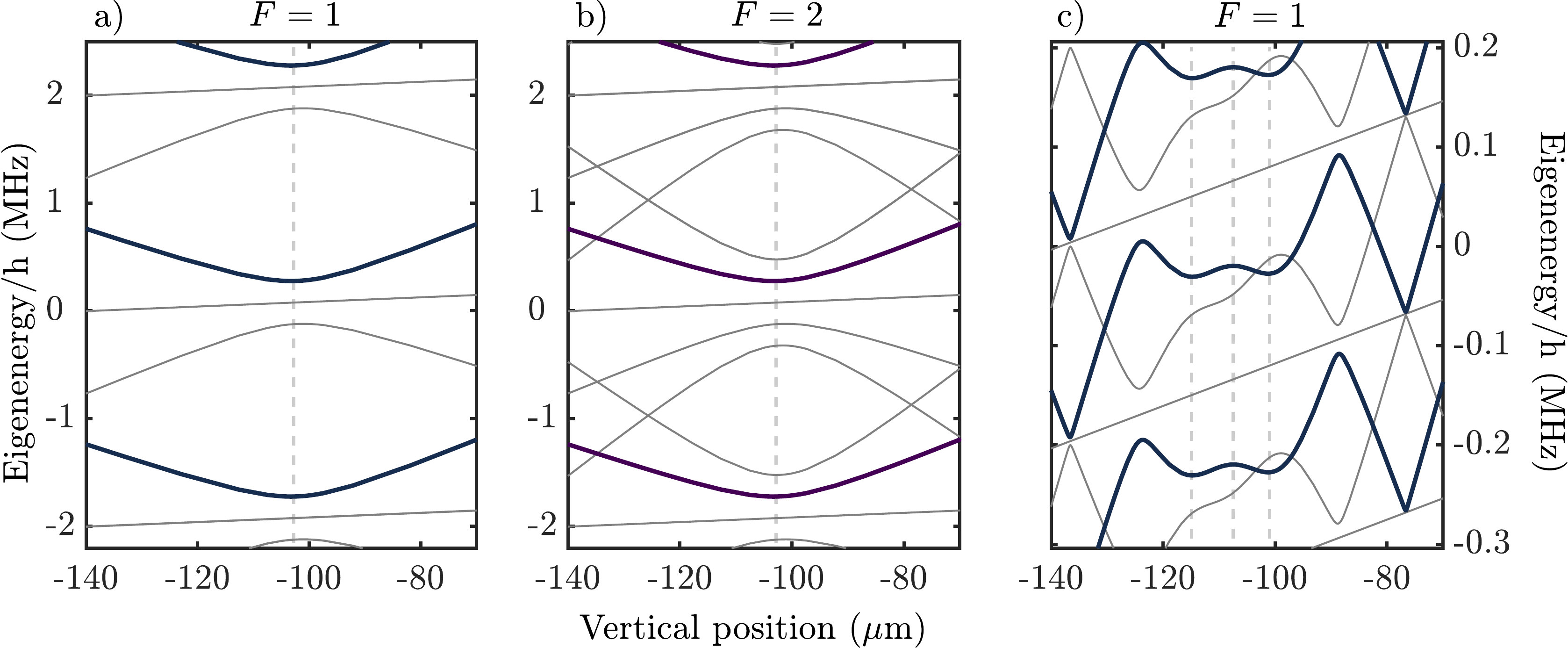}
	\caption{
		Eigenenergies of the dressed states as a function of vertical \mbox{position $z$}. The eigenstates with \mFt{1} are emphasised in bold and the zero of energy is arbitrary.
		Grey lines indicate eigenenergies of other dressed states which are not discussed in this work.
		a) \& b) Eigenenergies of dressed states of \Rb{87} with \Fo{} and \Ft, respectively, when dressed by a single \rf{} field at \SI{2}{\mega\hertz} and amplitude \SI{570}{\milli\gauss}.
		c) Eigenenergies of \Rb{87} with \Fo{} when dressed with three \rf{} components at frequencies \SIlist{1.8;2.0;2.2}{\mega\Hz} with amplitudes \SIlist{330;240;440}{\milli\gauss}.
		A magnetic quadrupole field of gradient of $B'= \SI{139.5}{\gauss\per\centi\metre}$ is used for all plots.
		Dashed lines indicate the positions where the resonance condition (equation~\ref{eq:resonanceCondition}) for each \rf{} component is satisfied.} 
	\label{fig:introToAPs}
\end{figure}

In the rotating-wave approximation, and incorporating gravity, the eigenenergies of a dressed atom at position $\mathbf{r} = (x,y,z)$ take the form

\begin{equation}
\label{eq:dressedAtomEnergies}
U(\mathbf{r}) = \tilde{m}_F \hbar \sqrt{\delta(\mathbf{r})^2 + \Omega(\mathbf{r})^2} + \tilde{N} \hbar \omega + M g z \hspace{5pt} ,
\end{equation}

\noindent where $\tilde{m}_F$ labels each dressed eigenstate, $\Omega(\mathbf{r})$ is the Rabi frequency of the dressing field, $M$ is atomic mass, $g$ is gravitational acceleration, $z$ is vertical position and $\delta(\mathbf{r})=\omega-\gF{}\Bohrmag{}B/\hbar{}$ is the detuning between the applied \rf{} field with angular frequency $\omega$ and the Zeeman splitting of the atoms. 
The avoided crossings between eigenstates occur at spatial locations where the energy of an \rf{} photon is equal to the Zeeman splitting, fulfilling the resonance condition 
\begin{equation}
|\gF{}| \Bohrmag B(\mathbf{r}) = \hbar \omega \hspace{5pt} .
\label{eq:resonanceCondition}
\end{equation}
Figure~\ref{fig:introToAPs}~a) \& b) illustrate the dressed eigenenergies for the cases of a \Rb{87} atom with \Fo{} and \Ft, respectively, for a dressing field of frequency \SI{2.0}{\mega\hertz} and amplitude $2 \hbar \Omega / \gF \mu_B = \SI{570}{\milli\gauss}$, with a magnetic quadrupole field of gradient $B'= \SI{139.5}{\gauss\per\centi\metre}$.
States with \mFt{1} are labelled in bold and are discussed in section~\ref{sec:methods}.
The spatial variation of the eigenenergies with $\tilde{m}_F > 0$ creates a trapping potential~\cite{Hofferberth2007,Perrin2017} where atoms are trapped in states for which $U$ has a local minimum near $\delta(\mathbf{r}) = 0$, as indicated by the dashed lines in figure~\ref{fig:introToAPs}. 
For atoms in the static quadrupole field of \refeqn{eq:quadrupole}, the resonance condition 
(equation~\ref{eq:resonanceCondition}) is satisfied on the surface of an oblate spheroid centered on the origin, forming a `shell' on which atoms are confined. 
Under the influence of gravity, atoms collect around the lowest point on this shell.

More complex potentials, such as double-wells, can be engineered by increasing the number of dressing \rf{} frequencies, as shown in figure~\ref{fig:introToAPs}~c)~\cite{Harte2018}. 
In section~\ref{sec:doublewell}, we present the realisation of a species-selective double-well potential.

\section{Experimental methods}
\label{sec:methods}

\begin{figure}[h]
	\centering
	\includegraphics[width = 1\linewidth]{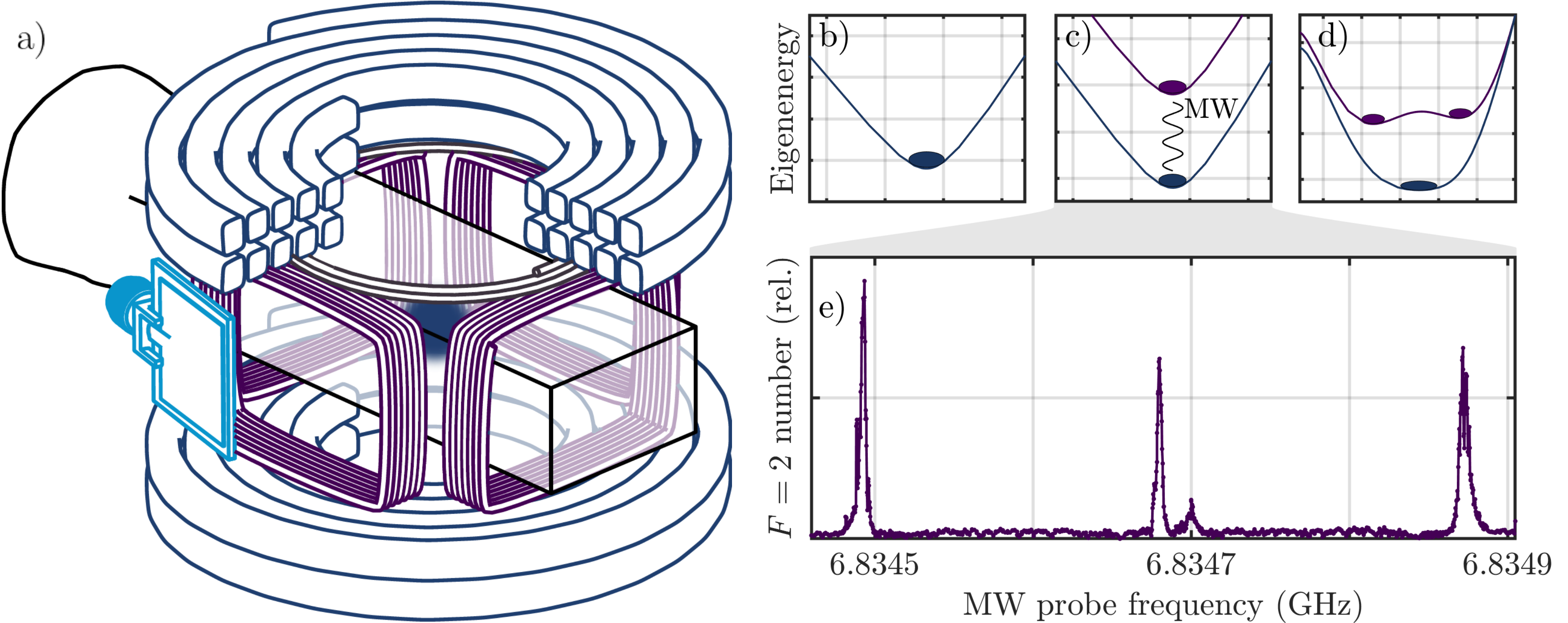}
	\caption{
		a) Schematic of the experimental apparatus, showing the magnetic quadrupole coils (dark blue), the \rf{}-dressing coil pairs (purple) which produce fields linearly polarised along $\mathbf{\hat{e}}_x$ and $\mathbf{\hat{e}}_y$, the \rf{} coil used for evaporation (grey), and the microwave patch antenna (light blue).
		The atoms (dark blue cloud) are confined within the ultra-high-vacuum glass cell (black), and a section of the upper quadrupole coil is cut away for clarity.
		\mbox{b)-d)} Illustration of an experimental sequence to produce an \rf-dressed mixture, where
		b) are the dressed eigenenergies of atoms in the \Fo{} state (dark blue);
		c) illustrates the transfer of a fraction of atoms into the \Ft{} state (purple) following a \mw{} pulse; and d) shows the potentials for each species as they are independently manipulated.
		e) Spectroscopic measurements of the \mw{} transitions between states with \Fo{}, \mFt{1} to untrapped \rf{}-dressed eigenstates with \Ft{}.}
	\label{fig:setup}
\end{figure}

Our experimental sequence follows that described in~\cite{Harte2018}. We produce cold atomic gases of \Rb{87} atoms in the state \Fo{}, \mFt{1}, which are trapped in a single-\rf-dressed potential and have been evaporatively cooled to approximately \SI{0.4}{\micro\kelvin}.
We implement our double-well potential by applying multiple \rf{} fields using current-carrying coil pairs, illustrated in figure~\ref{fig:setup}~a) (purple coils), which surround the ultra-high-vacuum glass cell.

To produce a mixture of hyperfine states, we apply a pulse of microwave (\mw{}) radiation using a patch antenna which is resonant with the hyperfine splitting of \Rb{87} at approximately \SI{6.8}{\giga\hertz}.
The \mw{} radiation is sourced from a commercial synthesizer\footnote{DS Instruments SG12000PRO} in series with a \SI{20}{\watt} amplifier\footnote{Microwave Amps, AM53-6.4-7-43-43}.
A pulse at 6.83468~\si{\giga\hertz} transfers a fraction of atoms from the \Fo, \mFt{1} state to the \Ft, \mFt{1} state.
The fraction of atoms transferred, and hence the relative densities of the two states, can be controlled via the \mw{} pulse duration.
We identify the relevant transition from a spectrum of \mw{} transitions, as shown in figure~\ref{fig:setup}~e), which is produced by applying a \SI{40}{\milli\second} \mw{} pulse at a given frequency and measuring the resulting atom number in states with \Ft{}. 
The features of this spectrum have been investigated extensively~\cite{Sinuco-Leon2019}.

We image both states in the same experimental sequence and, unless otherwise stated, while they are confined by the trapping potential (\textit{in situ}).
First, we image the \Ft{} cloud using light resonant with the $F=2$ to $F'=3$ transition in the D2 manifold (cooling light)~\cite{Steck2001}, where $F'$ denotes the angular momentum of the excited state. 
\Fo{} atoms are dark to this transition and are unaffected during this first image.
\Fo{} atoms are then optically pumped to $F=2$ using light resonant with the $F=1$ to $F'=2$ transition in the D2 manifold (repumping light) immediately before they are imaged with cooling light.
The two images are taken approximately \SI{6}{\milli\second} apart.

\section{A species-selective double-well potential}
\label{sec:doublewell}

We now describe our species-selective double-well potential, which we demonstrate for a mixture of the \Rb{87} hyperfine ground states with \Fo{} and \Ft{}.
We express the MRF dressing field as

\begin{equation}
\label{eq:field}
\begin{aligned}
\mathbf{B}_{\mathrm{\rf}} &= \sum_{i=1}^{N} B_i \Big(\cos\left(\omega_i t\right) \mathbf{\hat{e}}_x - \sin\left(\omega_i t\right) \mathbf{\hat{e}}_y \Big) + A_i \Big(\cos\left(\omega_i t\right) \mathbf{\hat{e}}_x + \sin\left(\omega_i t\right) \mathbf{\hat{e}}_y \Big) \\
\hspace{-10pt} &= \sum_{i=1}^{N}  \Big( B_i + A_i \Big) \cos\left(\omega_i t\right) \mathbf{\hat{e}}_x - \Big( B_i - A_i \Big) \sin\left(\omega_i t\right) \mathbf{\hat{e}}_y \hspace{5pt} ,
\end{aligned}
\end{equation}

\noindent with frequency components $i=1,2,...N$, at frequencies $\omega_i$.
$A_i$ and $B_i$ correspond to the circularly polarised \rf{} field amplitudes that couple states with \Fo{} and \Ft{}, respectively.
The sum and difference of $A_i$ and $B_i$ correspond to the magnitudes of fields linearly polarised along $\mathbf{\hat{e}}_x$ and $\mathbf{\hat{e}}_y$, respectively, which are produced by current-carrying coils, as illustrated in figure~\ref{fig:setup}~a). 
 
To create our double-well potential, we apply 3 \rf{} components with frequencies $\omega_i / 2\pi$ = \{\SIlist{1.8;2;2.2}{\mega\hertz}\} and amplitudes $(A_i + B_i)$ = \{\SIlist{330;370;440}{\milli\gauss}\}, which are linearly polarised along $\mathbf{\hat{e}}_x$. 
Each \rf{} field component $i$ creates an avoided crossing near the position $\mathbf{r}_i$ where the species-dependent resonance condition (equation~\ref{eq:resonanceCondition}) is satisfied, as shown in figure~\ref{fig:introToAPs}~c); the \rf{} field at \SI{2}{\mega\hertz} thus defines the `barrier' of the double-well potential.
Changing the polarisation of this field enables the barrier height to be independently controlled for each species, for instance, to raise the barrier for one species while simultaneously lowering it for the other.
We achieve this by the addition of an \rf{} field component at \SI{2}{\mega\hertz}, linearly polarised along $\mathbf{\hat{e}}_y$, and with amplitude $(B_2 - A_2)$.

\begin{figure}
	\centering
	\includegraphics[width=1\linewidth]{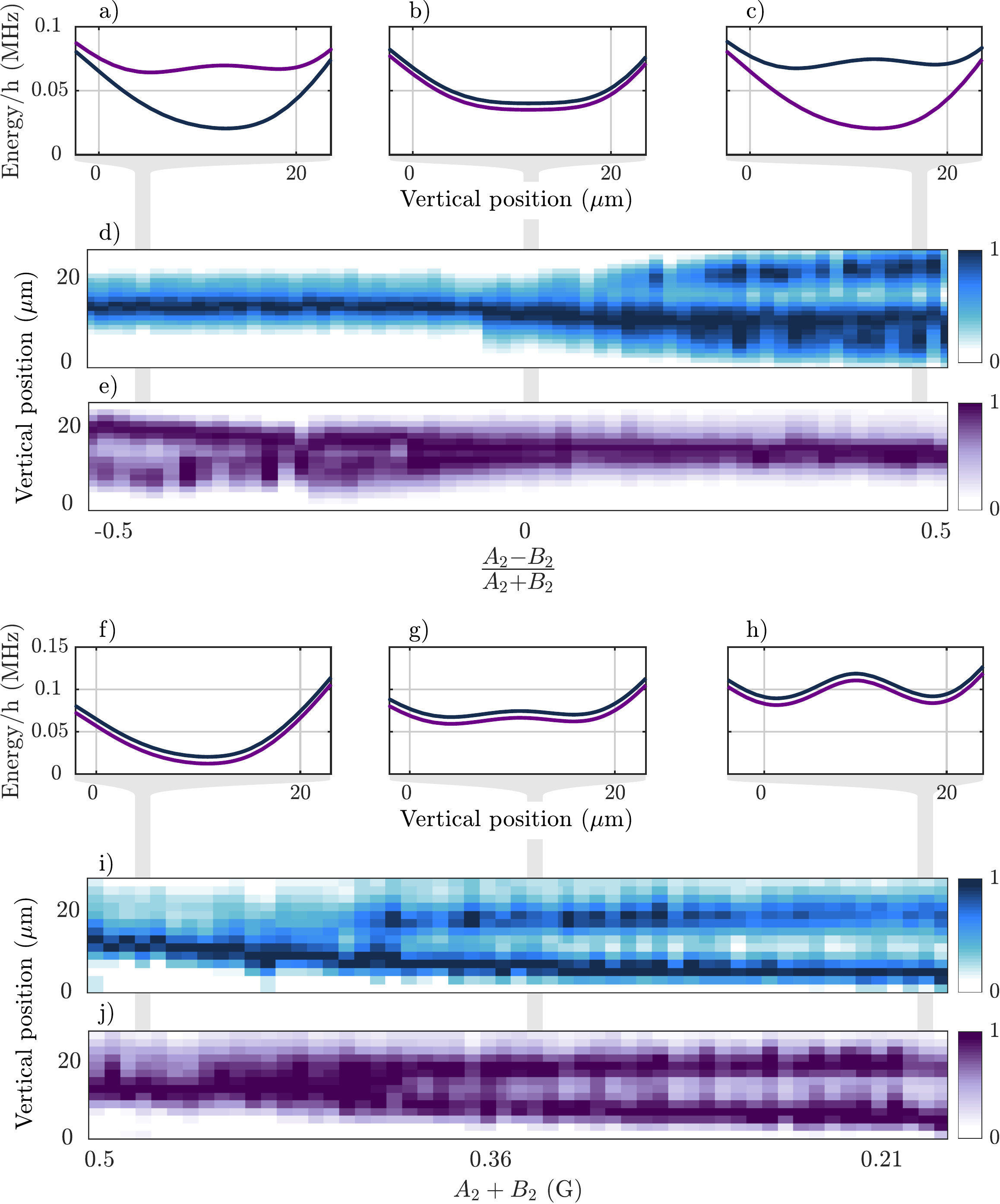}
	\caption{A species-selective double-well potential.
		a)-c) Potential energies for atoms with \Fo, \mFt{1} (blue) and \Ft, \mFt{1} (purple) for specific values of $(A_{2}-B_{2})/(A_{2}+B_{2})$.
		d)-e) \textit{In situ} density distributions of atoms with \Fo, \mFt{1} and \Ft, \mFt{1} vs.~$(A_{2}-B_{2})/(A_{2}+B_{2})$, respectively. The colour axes (right) indicate the relative number density. 
		f)-h) Potential energies of atoms with \Fo, \mFt{1} (blue) and \Ft, \mFt{1} (purple) for specific values of~$(A_{2}+B_2)$, where the dressing \rf{} field responsible for the barrier of the potential is linearly polarised along $\mathbf{\hat{e}}_x$.
		i)-j) \textit{In situ} density distributions of atoms with \Fo, \mFt{1} and \Ft, \mFt{1} vs.~$(A_{2}+B_2)$ respectively. 
		For clarity, an offset has been added to the eigenenergies of \Fo{} in b) and f)-h) to separate the \Fo{} and \Ft{} eigenenergies.
		A quadrupole gradient of \SI{136}{\gauss\per\centi\metre} is used for all plots.}
	\label{fig:Stripline}
\end{figure}

Figure~\ref{fig:Stripline}~a)-c) show the deformation of the potentials as the polarisation is changed. 
Changing the balance of $(A_{2}+B_{2})$ and $(A_{2}-B_{2})$ alters the amplitude of the $\sigma^{+}$ and $\sigma^{-}$ components, independently modifying the potential energies for \Fo{} and \Ft{}.
Figure~\ref{fig:Stripline}~d) \& e) show vertical slices of absorption images of the trapped \Fo{} and \Ft{} clouds, respectively, as the barrier polarisation is controlled.
Both species display the expected spatial distribution as the central \rf{} component is controlled, with the separation in position clearly illustrated.
Amplitude ramps are performed over timescales of order \SI{100}{\milli\second}, which is slower than the characteristic oscillation periods of atoms in the potentials.
This ensures the consequent deformation of the potentials does not excite unwanted motion in the trapped gases.

When no \rf{} field component at the frequency of the barrier is applied along $\mathbf{\hat{e}}_y$, the polarisation is linear along $\mathbf{\hat{e}}_x$ and the coupling strengths for \Fo{} and \Ft{} states are equal. 
As a consequence, raising or lowering the barrier via the amplitude of the \SI{2}{\mega\hertz} \rf{} field modifies the eigenstates equally for both species.
Figure~\ref{fig:Stripline}~f)-h) show the deformation of the potentials as the barrier field amplitude is reduced from  \SI{500}{\milli\gauss} to \SI{180}{\milli\gauss}. 
Figure~\ref{fig:Stripline} i) \& j) show vertical slices of absorption images of the trapped \Fo{} and \Ft{} clouds, respectively, as a function of the barrier amplitude $A_2+B_{2}$, which demonstrates the equivalence of potentials for both species.

In this section we have demonstrated a species-selective potential which is engineered via the amplitude and polarisation of dressing \rf{} fields. 
In particular, we demonstrated combinations of single- and double-well potentials for the two mixture constituents.
Furthermore, these methods are applicable to any mixture for which the magnetic nature of the constituents, as described by the Landé g-factors, is distinct.

\section{Collisional stability of the RF-dressed mixture}
\label{sec:collisions}

The species-selective manipulations we have demonstrated are only of practical use if the mixture is stable.
Recent work concerning a mixture of \Rb{85} and \Rb{87} indicates that most \rf{}-dressed mixtures are expected to undergo fast inelastic loss~\cite{Bentine2019a}; the atom-photon coupling allows inelastic collisions to conserve angular momentum through absorption or emission of an \rf{} photon, allowing spin exchange to occur with large rate coefficients~\cite{Owens2017,Bentine2019a}.
Fortunately, for collisions of \Rb{87} atoms, the rate coefficients for spin exchange are small, and it proceeds slowly even for cases that are allowed by angular momentum conservation rules. 
This special property of this particular isotope has previously been credited to the similarity of the singlet and triplet scattering lengths~\cite{Julienne1997}.
We note that mixtures of \Rb{87} hyperfine states with \Fo{} and \Ft{} have been used for interferometry in an \rf{}-dressed potential~\cite{Mas2019,Navez2016,Stevenson2015}.
Inelastic collisions of this mixture of \rf{}-dressed atoms have also been investigated in the context of \rf{}-dressed Feshbach resonances~\cite{Tscherbul2010}, however there has been no experimental study of the dependence of inelastic loss processes on the strength of the dressing \rf{} field.
We thus proceed to investigate the collisional stability of a mixture of \Rb{87} with \Fo{} and \Ft{} when confined in an \rf{}-dressed potential.

\subsection{F=1 alone}
\label{sec:F1}

We first consider a pure cloud of atoms in the \Fo{} state, where the atom number $N_1$ evolves in time as
\begin{equation}
\frac{\partial N_1}{\partial t} = - \alpha N_1 - \int k_2^{1,1} n_1^2 ~\mathrm{d}V - \int k_3^{1,1,1} n_1^3 ~\mathrm{d}V \hspace{5pt} ,
\label{eq:N1}
\end{equation}
where $\alpha$ is one-body rate coefficient, $n_1$ is the number density, and $k_2^{1,1}$ and $k_3^{1,1,1}$ are the two-body and three-body rate coefficients for of atoms with \Fo{}, respectively.
When inelastic loss is negligible, however, the terms with $k_2$ and $k_3$ in equation~\ref{eq:N1} can be discarded, and the atom number $N_1$ is observed to decay exponentially, at a rate given by $\alpha$. 

To measure the decay, we prepare a cloud of approximately $1.1 \times 10^5$ \Rb{87} atoms with \Fo{}, at a temperature of \SI{0.4}{\micro\kelvin}, in an \rf{}-dressed potential formed by a \SI{2}{\mega\hertz} \rf{} field and a quadrupole gradient of \SI{139.5}{\gauss\per\centi\metre}.
For these measurements, as well as those detailed later in section~\ref{sec:F12}, we image the gases after \SI{15}{\milli\second} of time-of-flight expansion after holding the atoms in the potential for a variable duration.
Figure~\ref{fig:Lifetimes}~a) (inset) illustrates the time-dependence of $N_1$ for three dressing \rf{} amplitudes (\SIlist{290;570;940}{\milli\gauss}).
We observe an exponential decay of $N_1$ in all cases, which consequently shows that inelastic loss is negligible for our range of densities, where the maximum number density across all samples is \SI{1.6d12}{\per\cm\cubed}.
Furthermore, by fitting the data with simplified versions of equation~\ref{eq:N1}, which contain only the two-body or three-body terms, respectively, we place bounds on the rate coefficients of $k_2 < \SI{1.6d-14}{\cm\cubed\per\s}$ and $k_3 < \SI{7.7d-27}{\cm^6~\s^{-1}}$.
An imperfect vacuum in the chamber and technical noise in the apparatus are the dominant causes of the observed decay for this single species.

\subsection{A mixture of F=1 and F=2}
\label{sec:F12}

\begin{figure}
	\centering
	\includegraphics[width = 0.95\linewidth]{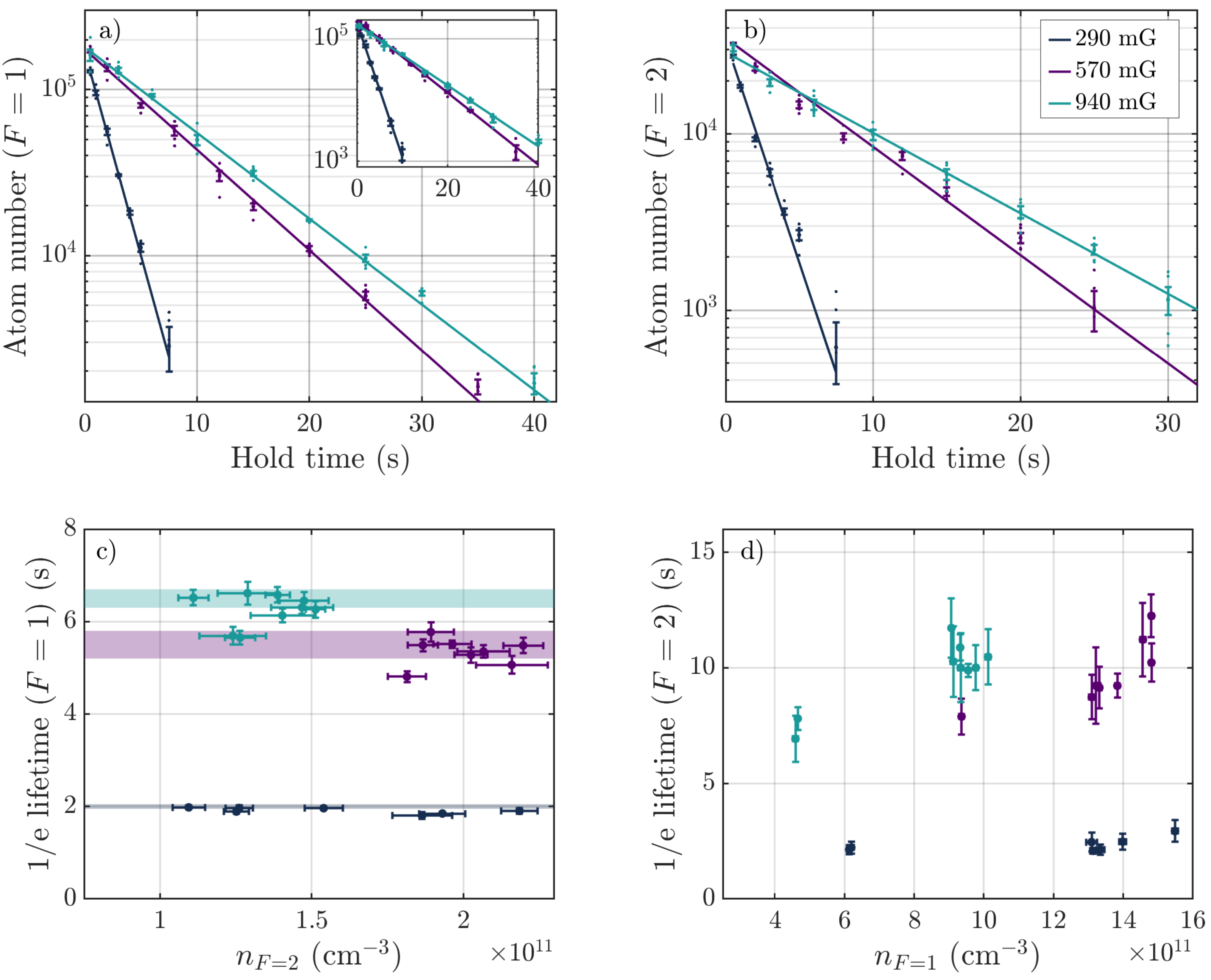}
	\caption{
		Decay of the number of atoms with \Fo{}, \mFt{1} and \Ft{}, \mFt{1} in an \rf-dressed mixture.
		Results are shown for clouds held in a single-frequency, linearly polarised \rf{}-dressed potential, for dressing \rf{} amplitudes of \SI{290}{\milli\gauss} (\textcolor[rgb]{0.0859,0.1758,0.3125}{\rule{7pt}{1.5pt}}), \SI{570}{\milli\gauss} (\textcolor[rgb]{0.4144,0,0.5179}{\rule{7pt}{1.5pt}}) and \SI{940}{\milli\gauss} (\textcolor[rgb]{0.1,0.6,0.6}{\rule{7pt}{1.5pt}}). 
		a) $N_1$ vs. hold time for a mixture of atoms with \Fo{} and \Ft, as described in section~\ref{sec:F12}.
		The lines indicate fitted curves which represent exponential decay.
		Inset: exponential fits to $N_1$ vs. hold time in the absence of atoms with \Ft, for three dressing \rf{} amplitudes, as discussed in section~\ref{sec:F1}.
		b) $N_2$ vs. hold time for a mixture of atoms with \Fo{} and \Ft.
		Individual measurements in a) \& b) are illustrated by points, and vertical error bars indicate the standard error, centred about the mean, for 6 repeated measurements at each hold time. 
		c) Measured 1/e lifetimes for \Fo{} as a function of the initial number density of \Ft.
		Shaded regions represent the lifetimes of \Fo{} atoms alone when trapped in an identical potential, which are extracted from the fitted curves in figure~\ref{fig:Lifetimes}~a), (inset).
		d) Measured 1/e lifetimes for \Ft{} as a function of the initial number density of \Fo.
		The vertical error bars in c) \& d) correspond to the uncertainty in the fitted rate coefficient, while horizontal error bars indicate the uncertainty in the initial number density, of which some are narrower than the width of data points.}
	\label{fig:Lifetimes}
\end{figure}

We now consider a two-species mixture of atoms in dressed states with \Fo{}, \mFt{1} and \Ft{}, \mFt{1}.
This choice of dressed states ensures both clouds have approximately the same gravitational sag in the potential and thus spatially overlap~\cite{Bentine2017}.\footnote{There exists a small difference in $|\gF|$ between these species and, consequently, there is a minor separation between the spatial locations of the potential minima~\cite{Mas2019}. In our context, however, this separation is negligible compared to the spatial extent of the trapped clouds.}
The number of atoms with \Fo{} in the mixture evolves as
\begin{equation}
\begin{aligned}
\frac{\partial N_1}{\partial t} &= - \alpha N_1 - \underbrace{\int k_2^{1,1} n_1^2 ~\mathrm{d}V - \int k_3^{1,1,1} n_1^3 ~\mathrm{d}V}_\text{Negligible} \\
&- \int k_2^{1,2} n_1 n_2 ~\mathrm{d}V - \int k_3^{1,1,2} n_1^2 n_2 ~\mathrm{d}V - \int k_3^{1,2,2} n_1 n_2^2 ~\mathrm{d}V \hspace{5pt} .
\end{aligned}
\label{eq:N12}
\end{equation}
where $n_2$ is the number density of atoms with \Ft{}, and $k_2^{i,j}$ and $k_3^{i,j,k}$ are the rate coefficients for two-body and three-body inelastic collisions of atoms in states $\{i,j,k\} \in \{1,2\}$, where $\{1,2\}$ represent the dressed states with \Fo{} and \Ft{}, respectively.
Measurements of \Fo{} alone (figure~\ref{fig:Lifetimes}~a) (inset)) have shown terms containing $k_1^{1,1}$ and $k_1^{1,1,1}$ to be negligible, which is indicated in equation~\ref{eq:N12}.

We experimentally verify the time-dependence of $N_1$ in the presence of atoms with \Ft.
To measure the inter-species rate constants, $k_2^{1,2}$, $k_3^{1,1,2}$ and $k_3^{1,2,2}$, it is sufficient to consider either loss of a small number of \Fo{} atoms in a \Ft{} bath, or to reverse the role of the species.
We use the same experimental parameters as section~\ref{sec:F1}, and the initial densities of atom clouds with \Fo, \mFt{1} and \Ft, \mFt{1} are controlled via the \mw{} pulse duration, which determines the number of atoms $N_2$ which are transferred from the state with \Fo{}.
We thus vary the initial densities of atoms with \Fo{} and \Ft{} between $4-16 \times 10^{11}$cm$^{-3}$ and $1-2 \times 10^{11}$ cm$^{-3}$, respectively.

Figure~\ref{fig:Lifetimes}~a)~\&~b) show $N_1$ and $N_2$ as a function of hold time, respectively, for an \rf-dressed mixture.
As \rf{} field strength has been predicted to modify the inter-species rate coefficients~\cite{Owens2017}, we also repeat the measurements for three dressing field amplitudes (\SIlist{290;570;940}{\milli\gauss}).
The decay is clearly exponential for all measurements: for both $F$ and for all three field strengths.
This implies that inelastic loss remains negligible for these species over our range of densities.
Again, by simplifying equation~\ref{eq:N12} to include only the terms with $k_2^{1,2}$, and either $k_2^{1,2,2}$ or $k_3^{1,1,2}$, we determine the upper bounds of $k_2^{1,2} < \SI{2.2d-14}{\cm\cubed~\s^{-1}}$ and $k_3~ (\mathrm{all~mechanisms})< \SI{1.2d-25}{\cm^6~\s^{-1}}$.

To further illustrate the lack of density dependence in the observed time-dependent dynamics, figure~\ref{fig:Lifetimes}~c) shows the 1/$e$ lifetimes for \Fo{}, which are extracted from the fitted constants 1/$\alpha$, for each dressing \rf{} field amplitude and over a range of initial $n_2$.
Coloured boxes illustrate the 1/$e$ lifetimes (and uncertainties) for the \Fo{} population in the absence of a coexisting \Ft{} cloud (figure~\ref{fig:Lifetimes}~a) (inset)). 
A surprising feature is that the lifetime of \Ft{} atoms vs. $n_1$, as shown in figure~\ref{fig:Lifetimes}~d), rises as $n_1$ is increased.
We speculate this is due to a temperature-dependence of the lifetime $1/\alpha$; longer microwave pulses are required to transfer more atoms from \Fo{} to \Ft{} and these pulses induce heating in the sample.
Temperature dependence in the lifetimes of trapped gases has been observed elsewhere~\cite{Rem2013,Fletcher2013}, albeit observations in these case were attributed to two- and three-body effects.
As we begin with \Fo{} and transfer a fraction to \Ft{}, rate coefficients for \Ft{} alone in the \rf{}-dressed potential are not determined. 
Additionally, in figure~\ref{fig:Lifetimes}, the lifetimes and initial densities of atoms with \Ft{} have a larger standard error than those for \Fo{}; this occurs because the number of atoms with \Ft{} is small compared to those with \Fo{}, even at maximum, as the imperfect \mw{} transfer leads to atom loss and heating.

\section{Conclusion and Outlook}

We have experimentally realised species-selective manipulations of an atomic mixture using MRF-dressed potentials, which are made possible by the different \gF{} of the constituents.
We have demonstrated that combinations of single- and double-well potentials for mixture constituents are made possible through control of the polarisations and amplitudes of dressing \rf{} fields.
Furthermore, the scales of spatial variation can be easily tuned by the choice of dressing field frequencies and magnetic field gradient. 
In a previous experiment, MRF-dressed potentials were used to realise a double-well potential with well separations of tens of \si{\micro\metre}~\cite{Harte2018}, and we have since reduced this separation to 2.5 \si{\micro\metre}~\cite{Barker2020a}.

We do not observe inelastic loss for a mixture of \Rb{87} atoms in \rf{}-dressed states with \Fo{} and \Ft{} over the range of densities and dressing field strengths used in these experiments.
The long lifetimes of \rf{}-dressed mixtures of \Rb{87} hyperfine states make this mixture highly promising for future experiments, such as probing impurity physics or non-equilibrium dynamics. 
In addition, the lifetime remains favourable for clouds of higher atomic number density. 
A preliminary measurement has shown that the lifetime of \Fo{} and \Ft{} BECs is of order seconds, although immiscibility may reduce the effective overlap of the species.

\section*{Acknowledgements}
This work was supported by the EPSRC Grant Reference EP/S013105/1. We gratefully acknowledge the support of NVIDIA Corporation with the donation of the Titan Xp GPU used for parts of this research. AJB, KL, DG and AB thank the EPSRC for doctoral training funding.

\section*{References}

\appendix

\bibliographystyle{iopart-num}
\bibliography{library}

\end{document}